\documentclass[prd,showpacs,twocolumn,superscriptaddress,floatfix,nofootinbib,10pt]{revtex4-2}
\usepackage{setspace}
\usepackage[utf8]{inputenc}
\usepackage{float}
\usepackage{amsmath,amssymb,amsfonts,bm}
\usepackage{graphicx}
\usepackage{multirow}
\usepackage[usenames,dvipsnames]{color}
\allowdisplaybreaks
\usepackage[
colorlinks=true,
linkcolor=blue,
breaklinks=true,
urlcolor=blue,
citecolor=blue]{hyperref}

\usepackage{orcidlink}
\usepackage{subfigure}
\usepackage{soul}
\usepackage{xcolor}
\usepackage{physics}
\usepackage{booktabs}
\usepackage{ulem}

\definecolor{green}{rgb}{0,0.6,0}
\newcommand{\mev}{\textrm{ MeV}}

\newcommand{\GXNU}{\affiliation{Department of Physics, Guangxi Normal University, Guilin 541004, China}}

\newcommand{\GXZD}{\affiliation{Guangxi Key Laboratory of Nuclear Physics and Technology, Guangxi Normal University, Guilin 541004, China}}

\newcommand{\BeiHangU}{\affiliation{School of Physics, Beihang University, Beijing, 102206, China}}

\newcommand{\HuNanNU}{\affiliation{Department of Physics, Hunan Normal University, Changsha 410081, China}}

\newcommand{\IFIC}{\affiliation{Departamento de F\'{\i}sica Te\'orica and IFIC, Centro Mixto Universidad de
Valencia-CSIC Institutos de Investigaci\'on de Paterna, Apartado 22085,
46071 Valencia, Spain}}

\begin{document}
\title{Theoretical description of the $D^{+} \to \bar{K}^{0} \pi^{+} \pi^{0} \pi^{0}$ reaction}

\begin{abstract}
We study the $D^{+} \rightarrow \bar{K}^{0} \pi^{+} \pi^{0} \pi^{0}$ reaction, which shows two clear structures: the $\bar{K}^{*0}$ and $\rho^{+}$ in the $\bar{K}^{0} \pi^{0}$ and $\pi^{+} \pi^{0}$ mass distributions, respectively. The study is done starting from the quark level with external and internal emission and hadronizing a pair of quarks to produce a vector and two pseudoscalars. We also consider another mechanism, which does not require hadronization of quark pairs: the direct production of $\bar{K}^{*0}$ and $\rho^{+}$. With the help of six unknown parameters which are fitted to the data, we are able to get a very good agreement with experimental data for the $\bar{K}^{0} \pi^{0}$ and $\pi^{+} \pi^{0}$ mass distributions and a qualitative one for the rest of mass distributions which do not show any particular structure experimentally.
\end{abstract} 

\author{Zi-Ying Yang}
\BeiHangU%

\author{Wen-Hao Jia}
\GXNU%

\author{Dao-Chong Lin}
\GXNU%

\author{Wei-Hong Liang\orcidlink{0000-0001-5847-2498}}%
\email{liangwh@gxnu.edu.cn}
\GXNU%
\GXZD%

\author{Chu-Wen Xiao}
\GXNU%
\GXZD%

\author{Raquel Molina}
\IFIC%

\author{Qi-Fang L\"u}
\HuNanNU%

\author{Eulogio Oset\orcidlink{0000-0002-4462-7919}}%
\email{Oset@ific.uv.es}
\GXNU%
\IFIC%

\maketitle

\section{Introduction}\label{sec:Intr}
Reactions with four bodies in the final state are common in collisions and particles decays \cite{BESIII:2025eqp,BESIII:2024ziy,LHCb:2022orj,BESIII:2017jyh,LHCb:2016qbq,ANKE:2007dyc}. 
Theoretical works on these reactions are not so abundant,  given the complexity required in the interpretation of the many invariant masses present in the reactions. Examples of the latter are the work of Ref.~\cite{Lebiedowicz:2009pj}, which studies the $pp \to pp \pi^{+} \pi^{-}$ reaction, and those of Refs.~\cite{Xie:2010md,Lu:2014rla}, which study the $pp \to pK^{+} K^{-} p$ and $pp \to pK^{+} \pi^{0} \Sigma^{0}$ reactions. 
Another example is the analysis of the $D_{s}^{+} \to \pi^{+} \pi^{+} \pi^{-} \eta$ decay carried out in Ref.~\cite{Song:2022kac}. 
This latter reaction was complicated given the number of resonances that were produced, $a_{0}(980)$, $f_{0}(980)$, $a_{1}(1260)$, and $b_{1}(1235)$, all of which were dynamically generated by the interaction of pairs of mesons, in addition to the $\rho$ meson, which is considered a standard $q \bar{q}$ state. 
Similar to that is the reaction that we study in the present work, the $D^{+} \to \bar{K}^{0} \pi^{+} \pi^{0} \pi^{0}$. 
However, even though they look similar, the dynamics of the two reactions are quite different. We will show that, in the present case, practically all the strength of the interaction goes into the production of $\bar{K}^{0}$ and $\rho^{+}$. 
Yet, there are subtleties in how these resonances are produced in the weak decays and with which weight, and also whether they are produced one at a time or the two of them simultaneously. 
In addition, the two identical $\pi^{0}$ particles impose constraints on the production amplitudes, with clear repercussions in the invariant mass distributions. 
We face all these issues and produce theoretical results that we compare with the experimental mass distributions of Ref.~\cite{BESIII:2023qgj} to draw conclusions on the dynamical mechanisms involved in the reaction.

\section{Formalism}\label{Sec:Formalism}
\subsection{External emission with one hadronization}
In the $D^{+} \to \bar{K}^{0} \pi^{+} \pi^{0} \pi^{0}$ decay of the BESIII experiment \cite{BESIII:2023qgj}, one can observe two clear structures, corresponding to the $\bar{K}^{*0}$ excitation in the $\bar{K}^{0} \pi^{0}$ invariant mass distribution, and the $\rho^{+}$, seen in the $\pi^{+} \pi^{0}$ mass distribution. 
In our formalism, we keep that in mind, and look at the production mechanisms where at least one of these two resonances is produced, together with the mechanism where the two resonances are simultaneously produced. 
To show how these mechanisms proceed, we investigate the production process at the quark level. 

\begin{figure}[t]
	\begin{center}
		\includegraphics[width=0.55\linewidth]{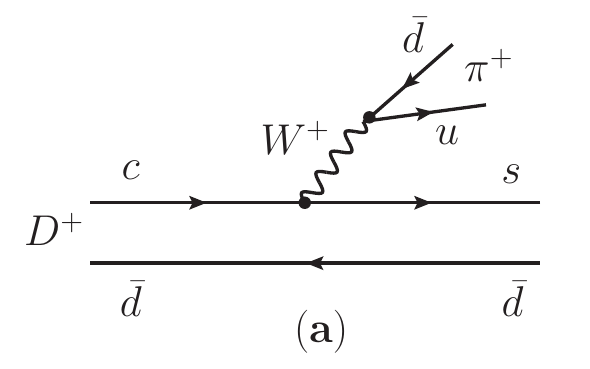}\\[2mm]
		\includegraphics[width=0.55\linewidth]{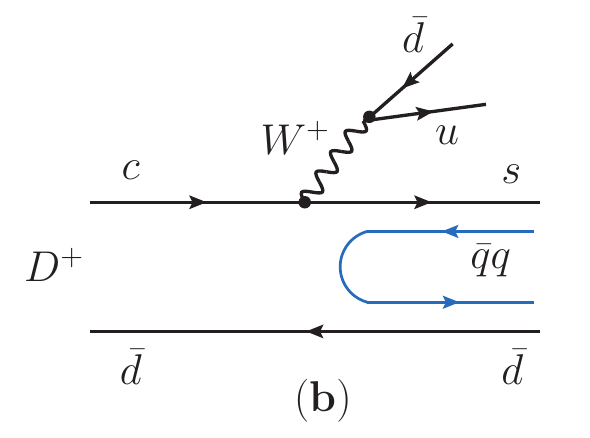}
		\vspace{-0.25cm}
		\caption{External emission mechanism for the decay of $D^{+}$ at the quark level: (a) basic mechanism; (b) hadronization of the $s\bar{d}$ component.}
		\label{fig:Fig1}
	\end{center}
\end{figure}

First, let us look into the large $N_{c}$ dominant external emission mode,
which is illustrated in Fig.~\ref{fig:Fig1} at the quark level.
To obtain the four daughter particles, one vector meson and two pseudoscalar mesons are involved, where the vector meson subsequently decays into two pseudoscalar mesons. 
We show this in Fig.~\ref{fig:Fig1}(a). 
To produce three final mesons, we must hadronize a pair of $q\bar{q}$ quarks. 
We start from the hadronization of the $s\bar{d}$ pair, as shown in Fig.~\ref{fig:Fig1}(b). 
We have two possibilities:
\begin{itemize}
	\item[(1)] The $u\bar{d}$ pair forms a $\pi^{+}$, in which case we have to hadronize the $s\bar{d}$ into a vector and a pseudoscalar;
	\item[(2)] The $u\bar{d}$ pair forms a $\rho^{+}$ and then the $s\bar{d}$ pair is hadronized into two pseudoscalars.
\end{itemize}

To achieve this hadronization, we represent $q\bar{q}$ in terms of mesons, as given by the following matrices:
\begin{align}
P &= \begin{pmatrix}
\frac{\pi^{0}}{\sqrt{2}} + \frac{\eta}{\sqrt{3}} + \frac{\eta^{\prime}}{\sqrt{6}} & \pi^{+} & K^{+}\\[1.5mm]
\pi^{-} & -\frac{\pi^{0}}{\sqrt{2}} + \frac{\eta}{\sqrt{3}} + \frac{\eta^{\prime}}{\sqrt{6}} & K^{0}\\[1.5mm]
K^{-} & \bar{K}^{0} & -\frac{\eta}{\sqrt{3}} + \frac{2\eta^{\prime}}{\sqrt{3}}
\end{pmatrix}, \label{eq:Pmatrix} \\
V &= \begin{pmatrix}
\frac{\rho^{0}}{\sqrt{2}} + \frac{\omega}{\sqrt{2}} & \rho^{+} & K^{*+}\\[1.5mm]
\rho^{-} & -\frac{\rho^{0}}{\sqrt{2}} + \frac{\omega}{\sqrt{2}} & K^{*0}\\[1.5mm]
K^{*-} & \bar{K}^{*0} & \phi
\end{pmatrix}, \label{eq:Vmatrix}
\end{align}
where the conventional $\eta-\eta^{\prime}$ mixing of Ref.~\cite{Bramon:1992kr} has been used. The results derived from the two options introduced above are analyzed below, omitting the $\eta^{\prime}$ contribution, which is irrelevant for these processes.

\begin{figure}[t]
	\begin{center}
		\includegraphics[width=0.62\linewidth]{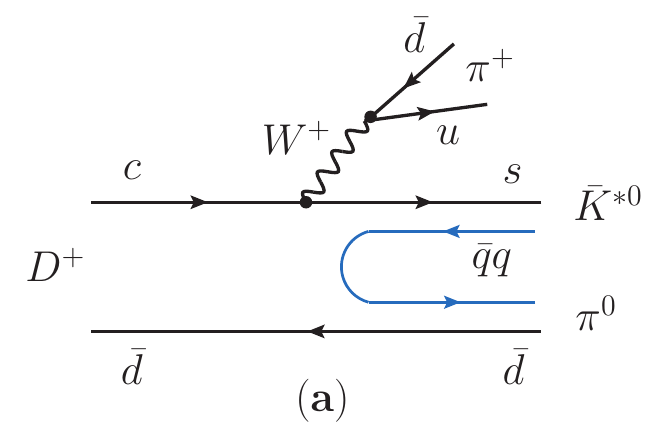}\\[1mm]
		\includegraphics[width=0.64\linewidth]{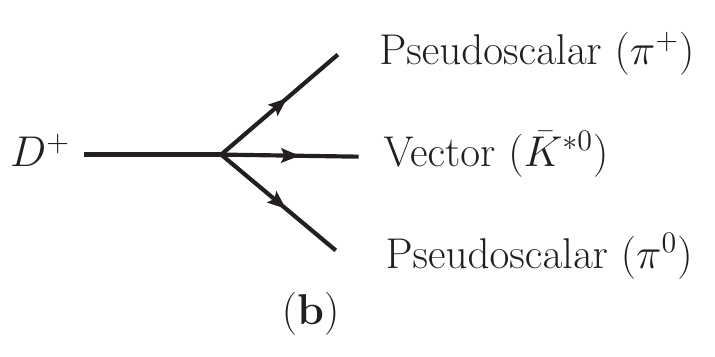}
		\vspace{-0.25cm}
		\caption{External emission mechanism for the $D^+ \to \pi^+ \bar K^{*0} \pi^0$ decay process: (a) quark-level diagram; (b) meson-level diagram. The $u\bar{d}$ pair forms a $\pi^{+}$ (pseudoscalar meson), while the remaining mesons are a vector meson and a pseudoscalar meson.}
		\label{fig:Fig2}
	\end{center}
\end{figure}

When the $u\bar{d}$ pair forms a $\pi^{+}$ we have:
\begin{itemize}
	\item[(1a)] Hadronization of the $s\bar{d}$ pair into a $PV$ (pseudoscalar meson and vector meson) configuration is performed, giving
       \begin{align}\label{eq:3}
         s\bar{d} \to & \sum_{i}s\bar{q}_{i}q_{i}\bar{d} = P_{3i}V_{i2} = (PV)_{32}\nonumber\\
         &= K^{-}\rho^{+} + \bar{K}^{0}\left(-\frac{\rho^{0}}{\sqrt{2}} + \frac{\omega}{\sqrt{2}}\right) - \frac{\eta}{\sqrt{3}}\bar{K}^{*0}.
       \end{align}
This process is not capable of producing the daughter particles $\bar{K}^{0}\pi^{+}\pi^{0}\pi^{0}$.
  \item[(1b)]When the $s\bar{d}$ pair is hadronized into a $VP$ (vector meson and pseudoscalar meson) configuration, we obtain
     \begin{align}\label{eq:4}
       s\bar{d}\to & \sum_{i}s\bar{q}_{i}q_{i}\bar{d} = V_{3i}P_{i2} = (VP)_{32} \nonumber\\
	   &= K^{*-}\pi^{+} + \bar{K}^{*0}\left(-\frac{\pi^{0}}{\sqrt{2}} + \frac{\eta}{\sqrt{3}}\right) + \phi \bar{K}^{0}. 
     \end{align}
In this case, the term $\bar{K}^{*0}\pi^{0}$ can lead to the desired final state through the $\bar{K}^{*0}\to \bar{K}^{0}\pi^{0}$ decay. The process above is shown in Fig.~\ref{fig:Fig2}.
\end{itemize}

\begin{figure}[t]
	\begin{center}
		\includegraphics[width=0.62\linewidth]{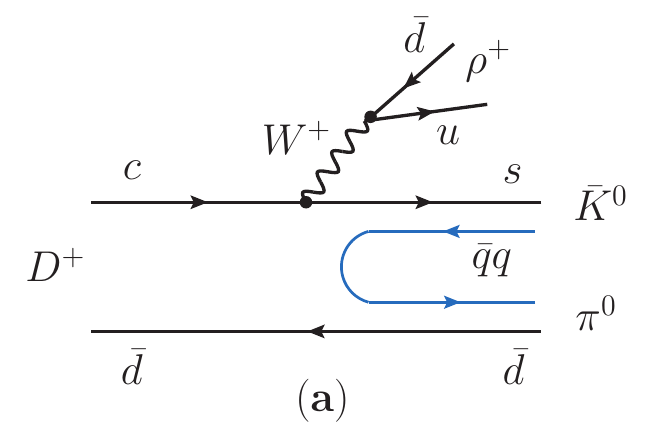}\\
		\includegraphics[width=0.64\linewidth]{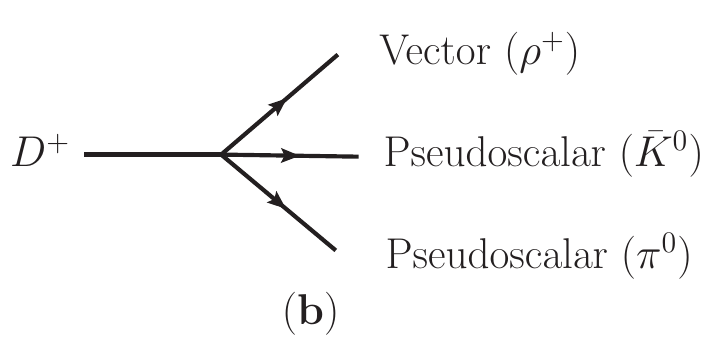}
		\vspace{-0.25cm}
		\caption{External emission mechanism for the $D^+ \to \rho^+ \bar K^0 \pi^0$decay process: (a) quark-level diagram; (b) meson-level diagram. The $u\bar{d}$ pair forms a $\rho^{+}$ (vector meson), while the remaining mesons are pseudoscalars.}
		\label{fig:Fig3}
	\end{center}
\end{figure}

Next we consider the second option, where the $u\bar{d}$ pair is now a $\rho^{+}$ and the $s\bar{d}$ pair is hadronized into a $PP$ (pseudoscalar-pseudoscalar meson) configuration, yielding
\begin{align}\label{eq:5}
s\bar{d}\to & \sum_{i}s\bar{q}_{i}q_{i}\bar{d} = P_{3i}P_{i2} = (P^{2})_{32} \nonumber\\
&= K^{-}\pi^{+} + \bar{K}^{0}\left(-\frac{\pi^{0}}{\sqrt{2}} + \frac{\eta}{\sqrt{3}}\right) - \frac{\eta}{\sqrt{3}}\bar{K}^{0}.
\end{align}
Here the $\bar{K}^{0}\pi^{0}$ term of Eq.~\eqref{eq:5}, in addition to the $\rho^{+}$ which decays to $\pi^{+}\pi^{0}$, again produces the desired final state $\bar{K}^{0}\pi^{+}\pi^{0}\pi^{0}$. 
This mechanism is depicted in Fig.~\ref{fig:Fig3}.

\begin{figure*}[t!]
\centering
\includegraphics[width=0.85\linewidth]{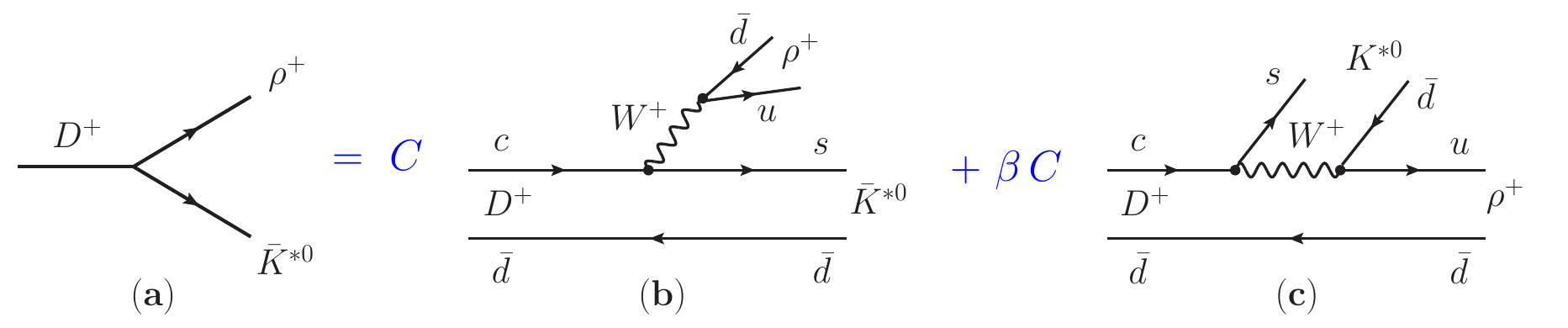}
\vspace{-0.25cm}
\caption{Mechanisms for $D^{+} \to \rho^{+} \bar{K}^{*0}$ decay: (a) meson-level representation; (b) quark-level diagram through external emission contributing with weight $C$; (c) quark-level diagram through internal emission contributing with weight $\beta C$.}
\label{fig:Fig4}
\end{figure*}   

\begin{figure*}[t]
\centering
\includegraphics[width=0.9\linewidth]{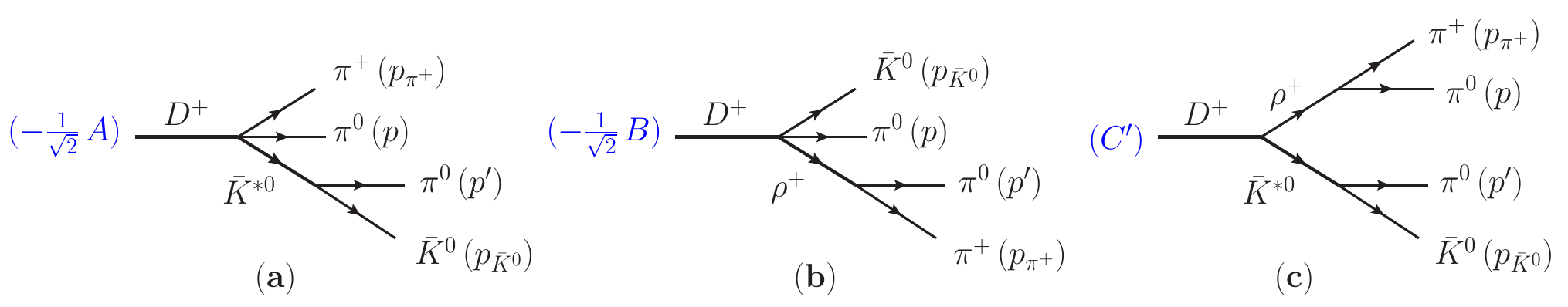}
\vspace{-0.25cm}
\caption{Diagrams of the decay mechanisms at the meson level, where the quantities in parentheses are the momenta of the mesons, 
and $-\frac{1}{\sqrt{2}}A$, $-\frac{1}{\sqrt{2}}B$, $C'$ (where $C' \equiv (1+\beta)C$) represent the vertex constants for $D^{+} \to \bar{K}^{*0}\pi^{+}\pi^{0}$, $D^{+} \to \bar{K}^{0}\pi^{0}\rho^{+}$, $D^{+} \to \rho^{+}\bar{K}^{*0}$, respectively. 
The factor $-\frac{1}{\sqrt{2}}$ stands for the weight of theses terms in Eqs.~\eqref{eq:4} and \eqref{eq:5}.
The overall diagrams includes this figure and the one obtained by exchanging $p$ and $p'$ (which is omitted for brevity).}
\label{fig:Fig5}
\end{figure*}

The third possibility, outlined alone, is given by the direct production of $\rho^{+}$ and $\bar{K}^{*0}$ which we describe below.

\subsection{Two-vector production without hadronization}

We have now two possibilities involving external and internal emission, shown in Fig.~\ref{fig:Fig4}.
In Fig.~\ref{fig:Fig4}(b), which proceeds via external emission, the $s$ quark and the $\bar{d}$ quark combine to form a $\bar{K}^{*0}$, while the $u$ and $\bar{d}$ quarks form a $\rho^{+}$. 
The $\bar{K}^{*0}$ subsequently decays into $\bar{K}^{0}\pi^{0}$, and the $\rho^{+}$ decays into $\pi^{+}\pi^{0}$, yielding the $\bar{K}^{0}\pi^{+}\pi^{0}\pi^{0}$ final state. 
In Fig.~\ref{fig:Fig4}(c), which proceeds through internal emission, the quark combination pattern is the same as in Fig.~\ref{fig:Fig4}(b).
We give weights $A$ to the terms of Eq.~\eqref{eq:4}, $B$ to the terms of Eq.~\eqref{eq:5} and $C$ to the terms from Fig.~\ref{fig:Fig4}(b).
Internal emission has a suppression factor of $\beta \simeq \frac{1}{N_{c}}$ with respect to external emission. 
If we give a weight $\beta C$ to internal emission, we can combine the two production mechanisms with a weight $C(1 + \beta) \equiv C'$.

One may wonder why we do not consider the hadronization of the $u\bar{d}$ component in Fig.~\ref{fig:Fig1}. 
First, if we hadronize the $u\bar d$ pair into two pseudoscalar mesons, there would be $\pi^{+}\pi^{0}$, but the vertex goes as $(\omega_{\pi^{+}} = \omega_{\pi^{0}})$ which cancels on average \cite{Sun:2015uva,Song:2025eko}.
If instead we have $s\bar{d} \equiv \bar{K}^{0}$ and hadronize the $u\bar{d}$ pair with $VP$ or $PV$ combinatorially, we can see that the needed $\rho^{+}\pi^{0}$ terms cancel in the sum of $VP + PV$.

We finalize by looking at the diagram of Fig.~\ref{fig:Fig4}(c) with internal emission at the quark level. 
If we hadronize the $s\bar{d}$ pair and the $u\bar{d}$ pair forms a $\rho^{+}$ meson, this gives a contribution to the $\rho^{+} \bar{K}^{0}\pi^{0}$ term obtained before from external emission. 
Similarly, we can have the $s\bar{d}$ pair to be a $\bar{K}^{0}$ and the $u\bar{d}$ pair hadronized into a pair of pions, which adds to the same term obtained earlier from external emission. 
We see that we do not get new structures but rather add contributions of order $\beta$ to the terms that we had before. 
We thus have the structure shown in Fig.~\ref{fig:Fig5}, that provides the amplitudes $t_a$, $t_b$, $t_c$, to which we give global weights including the external and internal production mechanisms of $A$, $B$, $C'$ respectively.

The analytical structure of these amplitudes is given in the next section.

\section{EVALUATION OF THE DIFFERENTIAL CROSS SECTION}

We follow closely the method used in Ref.~\cite{Song:2022kac}.

The width for the $D^{+}$ decay into $\bar{K}^{0}\pi^{+}\pi^{0}\pi^{0}$ is given by
\begin{align}\label{eq:6}
\Gamma &= \frac{1}{2M_{D^+}} \,\frac{1}{2}\int \frac{\dd^3 p_{\bar{K}^0}}{(2\pi)^3} \;\frac{1}{2\omega_{\bar{K}^0}} 
\int \frac{\dd^3 p_{\pi^+}}{(2\pi)^3} \;\frac{1}{2\omega_{\pi^+}}  \nonumber\\
& \times \int \frac{\dd^3 p_{\pi^0}}{(2\pi)^3} \;\frac{1}{2\omega_{\pi^0}} \int \frac{\dd^3 p'_{\pi^0}}{(2\pi)^3} \;\frac{1}{2\omega'_{\pi^0}} \nonumber\\
& \times (2\pi)^4 \delta^{(4)} (P - p_{\bar{K}^0} - p_{\pi^+} - p_{\pi^0} - p'_{\pi^0})\;|t|^2,
\end{align}
where $P$ is the $D^{+}$ momentum, the factor $\frac{1}{2}$ accounts for the identity of the two $\pi^{0}$, and $t$ is the amplitude for the $D^{+} \to \bar{K}^{0}\pi^{+}\pi^{0}\pi^{0}$ decay.

We eliminate the $p_{\pi^{+}}$ integration using the $\delta^{3}()$ function. 
Since we are working in the rest frame of the parent particle, which means that $\vec{P} = 0$:
\begin{equation}\label{eq:7}
\vec{p}_{\pi^{+}} = \vec{P} - \vec{p}_{\bar{K}^{0}} - \vec{p}_{\pi^{0}} - \vec{p}'_{\pi^{0}} = -(\vec{p}_{\bar{K}^{0}} + \vec{p}_{\pi^{0}} + \vec{p}\,'_{\pi^{0}}).
\end{equation}
Accordingly, the three-momentum delta removes the $\dd^{3}{p}_{\pi^{+}}$ integration, and we are left with the single energy-conserving delta.

For convenience, we define the variables $\vec{P}_{\pi}$ and $\vec{q}$ as the sum and difference of the two $\pi^{0}$ momenta, respectively:
\begin{equation}\label{eq:8}
\vec{P}_{\pi} \equiv  \vec{p}_{\pi^{0}} + \vec{p}\,'_{\pi^{0}}, \qquad \vec{q} \equiv \vec{p}_{\pi^{0}} - \vec{p}\,'_{\pi^{0}};  
\end{equation}
or equivalently:
\begin{equation}\label{eq:9}
\vec{p}_{\pi^{0}} = \frac{1}{2} (\vec{P}_{\pi} + \vec{q}\,), \qquad \vec{p}\,'_{\pi^{0}} = \frac{1}{2} (\vec{P}_{\pi} - \vec{q}\,). 
\end{equation}
Then Eq.~\eqref{eq:6} for $\Gamma$  is transformed into
\begin{align}\label{eq:10}
\Gamma = & \frac{1}{2M_{D^{+}}} \,\frac{1}{2} \int \frac{\dd^3p_{\bar{K}^0}}{(2\pi)^3} \frac{1}{2\omega_{\bar{K}^0}} 
 \int \frac{\dd^3p_{\pi^0}}{(2\pi)^3} \frac{1}{2\omega_{\pi^0}} 
\int \frac{\dd^3p'_{\pi^0}}{(2\pi)^3} \frac{1}{2\omega'_{\pi^0}} \nonumber\\
&\!\! \times \frac{1}{2\omega_{\pi^+}} (2\pi) \delta \left( M_D + -\omega_{\bar{K}^0} - \omega_{\pi^+} - \omega_{\pi^0} - \omega'_{\pi^0} \right) \;|t|^2,
\end{align}
with
\begin{equation}
\omega_{\pi^{+}} = \sqrt{m_{\pi^{+}}^{2} + (\vec{P}_{\pi} + \vec{p}_{\bar{K}^{0}})^{2}}.
\end{equation}

We perform the change of variables $(\vec{p}_{\pi^{0}},\vec{p}\,'_{\pi^{0}}) \to (\vec{P}_{\pi},\vec{q}\,)$, and calculate the corresponding Jacobian,
\begin{equation}
	\int f(\vec{p}_{\pi^{0}},\vec{p}\,'_{\pi^{0}}) \dd^{3}{p}_{\pi^{0}} \dd^{3}{p}\,'_{\pi^{0}} = \int f(\vec{P}_{\pi},\vec{q}\,) (\frac{1}{2})^{3} \dd^{3}{P}_{\pi} d^{3}{q}
\end{equation} 
therefore,
\begin{align}\label{eq:13}
\Gamma =& \frac{1}{2M_{D^{+}}} \,\frac{1}{2}  \left( \frac{1}{2} \right)^3 
\int \frac{\dd^3 p_{\bar{K}^0}}{(2\pi)^3} \frac{1}{2\omega_{\bar{K}^0}} 
\int \frac{\dd^3 P_{\pi}}{(2\pi)^3} \int \frac{\dd^3 q}{(2\pi)^3}   \nonumber \\
& \!\!\!\times \frac{1}{2\omega_{\pi^0}} \,\frac{1}{2\omega_{\pi^+}}\; 2\pi \;\delta \left( M_{D^+} -\omega_{\bar{K}^0} - \omega_{\pi^+} - \omega_{\pi^0} - \omega'_{\pi^0} \right)\; |t|^2.
\end{align}
The argument of the $\delta()$ depends on the angle $\theta$ between $\vec{P}_{\pi}$ and $\vec{p}_{\bar{K}^{0}}$. 
The $\cos \theta$ integration is used to kill the $\delta()$ function, and then $\cos \theta$ can be written in terms of the other variables. 
We find
\begin{align}\label{eq:14}
\cos \theta & \equiv \lambda = \frac{1}{2p_{\bar{K}^{0}}P_{\pi}}\nonumber\\
& \times\left[(M_{D^{+}} - \omega_{\bar{K}^{0}} - \omega_{\pi^{0}} - \omega'_{\pi^{0}})^{2} - m_{\pi^{+}}^{2} - \vec{p}_{\bar{K}^{0}}^{\,2} - \vec{P}_{\pi}^{2}\right]. 
\end{align}
We must ensure that $|\cos \theta| \leq 1$ and include this constraint as a factor in the integrand
\begin{equation}
\Theta (1 - \lambda^{2}) \;\Theta (M_{D^{+}} - \omega_{\bar{K}^{0}} - \omega_{\pi^{0}} - \omega'_{\pi^{0}}),
\end{equation}
and then we take
\begin{equation}
\vec{p}_{\bar{K}^{0}} = p_{\bar{K}^{0}} \begin{pmatrix} 0 \\ 0 \\ 1 \end{pmatrix}, 
\end{equation}
and due to rotational symmetry of the whole system we can write $\int \dd^{3}p_{\bar{K}^{0}} = 4\pi \int p_{\bar{K}^{0}}^{2} \dd p_{\bar{K}^{0}}$. 
Then using
\begin{equation} 
\int \dd \lambda  \, \delta(f(\lambda)) = \frac{1}{|f'(\lambda)|},
\end{equation} 
we obtain
\begin{align}
\Gamma & = \frac{1}{2M_{D^{+}}} \,\frac{1}{16}\frac{1}{2\pi} \int p_{\bar{K}^0} \,\dd p_{\bar{K}^0} \,\frac{1}{2\omega_{\bar{K}^0}} 
\int \frac{P_{\pi} \, \dd P_{\pi} \, \dd\phi}{(2\pi)^3} \nonumber \\
& \times \int \frac{\dd^3q}{(2\pi)^3} \frac{1}{2\omega_{\pi^0}} \frac{1}{2\omega'_{\pi^0}} \nonumber \\
& \times |t|^2 \Theta(1 - \lambda^2) \,\Theta(M_D^+ - \omega_{\bar{K}^0} - \omega_{\pi^0} - \omega'_{\pi^0}),
\end{align}
with the value of $\cos \theta$ obtained before, we write now
\begin{equation}
\vec{P}_{\pi} = P_{\pi} \begin{pmatrix} \sin \theta \cos \phi \\ \sin \theta \sin \phi \\ \cos \theta \end{pmatrix}, 
\end{equation}
with $\sin \theta = \sqrt{1 - \lambda^{2}}$.

For the integration, according to Eq.~\eqref{eq:9}, it is convenient to define $\vec{q}$ with respect to $\vec{P}_{\pi}$, i.e. $\tilde{\vec{q}}$, as if $\vec{P}_{\pi}$ were oriented along the $z$ axis,
\begin{equation}
\tilde{\vec{q}} = q \begin{pmatrix} \sin \tilde{\theta}_{q} \cos \tilde{\phi}_{q} \\ \sin \tilde{\theta}_{q} \sin \tilde{\phi}_{q} \\ \cos \tilde{\theta}_{q} \end{pmatrix}. 
\end{equation}
We perform two rotations and obtain $\vec{q} = R \,\tilde{\vec{q}}$, with
\begin{align}
R = R_{\phi} R_{\theta} &= \begin{pmatrix}
\cos \phi & -\sin \phi & 0 \\
\sin \phi & \cos \phi & 0 \\
0 & 0 & 1
\end{pmatrix}
\begin{pmatrix}
\cos \theta & 0 & \sin \theta \\
0 & 1 & 0 \\
-\sin \theta & 0 & \cos \theta
\end{pmatrix} \nonumber \\[2mm]
&= \begin{pmatrix}
\cos \phi \cos \theta & -\sin \phi & \cos \phi \sin \theta \\
\sin \phi \cos \theta & \cos \phi & \sin \phi \sin \theta \\
-\sin \theta & 0 & \cos \theta
\end{pmatrix}. 
\end{align}
To define $\cos \theta \equiv \lambda$ of Eq.~\eqref{eq:14}, we need $\omega_{\pi^{0}} + \omega'_{\pi^{0}}$,
\begin{align}
\omega_{\pi^{0}} + \omega'_{\pi^{0}} &= \sqrt{m_{\pi^{0}}^{2} + \left[\frac{1}{2} (\vec{P}_{\pi} + \vec{q}\,)\right]^{2}} \nonumber\\[1mm]
&~~+ \sqrt{m_{\pi^{0}}^{2} + \left[\frac{1}{2} (\vec{P}_{\pi} - \vec{q}\,)\right]^{2}} \nonumber \\[1mm]
&= \sqrt{m_{\pi^{0}}^{2} + \frac{1}{4} P_{\pi}^{2} + \frac{1}{2} P_{\pi} q \cos \tilde{\theta}_{q} + \frac{1}{4} q^{2}} \nonumber\\[1mm]
&~~+ \sqrt{m_{\pi^{0}}^{2} + \frac{1}{4} P_{\pi}^{2} - \frac{1}{2} P_{\pi} q \cos \tilde{\theta}_{q} + \frac{1}{4} q^{2}}. 
\end{align}

Then we take $\omega_{\bar{K}^{0}} \; \dd\omega_{\bar{K}^{0}} = p_{\bar{K}^{0}} \; \dd p_{\bar{K}^{0}}$ into $\Gamma$, and finally find
\begin{align}\label{eq:FinalGamma}
\Gamma &= \frac{1}{2M_{D^{+}}} \frac{1}{32} \frac{1}{2\pi} \int \dd\omega_{\bar{K}^{0}} \int \frac{P_{\pi} \, \dd P_{\pi} \, \dd \phi}{(2\pi)^{3}} \nonumber \\[1mm]
&\quad \times\int \frac{q^{2} \; \dd q \, \dd\cos\tilde{\theta}_{q} \, \dd\tilde{\phi}_{q}}{(2\pi)^{3}} \,\frac{1}{2\omega_{\pi^{0}}} \,\frac{1}{2\omega'_{\pi^{0}}} \nonumber \\[1mm]
&\quad \times |t|^{2} \;\Theta (1 - \lambda^{2}) \;\Theta (M_{D^{+}} - \omega_{\bar{K}^{0}} - \omega_{\pi^{0}} - \omega'_{\pi^{0}}). 
\end{align}
The amplitude $t$ in Eq.~\eqref{eq:FinalGamma} is given as
\begin{equation}
t = t_{a} + t_{b} + t_{c} + \tilde{t}_{a} + \tilde{t}_{b} + \tilde{t}_{c}, 
\end{equation}
with $t_{a}$, $t_{b}$ and $t_{c}$ the amplitudes corresponding to the three diagrams (a), (b) and (c) in Fig.~\ref{fig:Fig5}, and $\tilde{t}_{a}$, $\tilde{t}_{b}$ and $\tilde{t}_{c}$ the amplitudes for these diagrams obtained by exchanging $p$ and $p'$, which are omitted in Fig.~\ref{fig:Fig5} for brevity.

As shown in Fig.~\ref{fig:Fig5}, the expressions for $t_{a}$, $t_{b}$ and $t_{c}$ can be derived from these diagrams.
We must take into account that the $\bar K^{*0} \to \bar K^0 \pi^0$ decay has a vertex proportional to $(\vec p_{\bar K^{0}}- \vec p_{\pi^0}) \cdot \vec \epsilon_{K^{*0}}$. 
Similarly, for the $\rho^+\to \pi^+\pi^0$, the vertex goes as $(\vec p_{\pi^+}- \vec p_{\pi^0}) \cdot \vec \epsilon_{\rho^+}$.
We have another $K^*$ or $\rho$ vertex in Fig.~\ref{fig:Fig5}(a) and (b), where, for reasons of symmetry, we contract the polarization vector with the sum of the momenta of the two pseudoscalars not coming from the vector decay. 
We include the coupling constants for the vector decay and the original weights $-\frac{1}{\sqrt{2}} A$, $-\frac{1}{\sqrt{2}} B$, and $(1 + \beta) C$ of the diagrams in Fig.~\ref{fig:Fig5} into new coefficients $A'$, $B'$ and $C'$, 
and we obtain
\begin{align} 
t_{a} &= A'\, (\vec{p} + \vec{p}_{\pi^{+}})\cdot (\vec{p}_{\bar K^{0}} - \vec{p}\,') \nonumber\\
&~~~~\times\frac{1}{(p' + p_{\bar K^{0}})^{2} - M_{\bar K^{*0}}^{2} + i M_{\bar K^{*0}} \Gamma_{\bar K^{*0}}}, \\
t_{b} &= B'\, (\vec{p} + \vec{p}_{\bar K^{0}})\cdot (\vec{p}_{\pi^{+}} - \vec{p}\,') \nonumber\\
&~~~~\times\frac{1}{(p' + p_{\pi^{+}})^{2} - M_{\rho^{+}}^{2} + i M_{\rho^{+}} \Gamma_{\rho^{+}}}, \\
t_{c} &= C'\, M_{\bar K^{*0}}^{2} (\vec{p}_{\pi^{+}} - \vec{p}\,)\cdot (\vec{p}_{\bar K^{0}} - \vec{p}\,') \nonumber\\
&~~~~\times\frac{1}{(p_{\pi^{+}} + p)^{2} - M_{\rho^{+}}^{2} + i M_{\rho^{+}} \Gamma_{\rho^{+}}} \nonumber\\
&~~~~\times \frac{1}{(p' + p_{\bar K^{0}})^{2} - M_{\bar K^{*0}}^{2} + i M_{\bar K^{*0}} \Gamma_{\bar K^{*0}}}, 
\end{align}
where we introduce the factors $M_{\bar K^{*0}}^{2}$ in $t_c$ to have the coefficients $A'$, $B'$, $C'$ with the same dimension.

We must introduce now the terms symmetrizing the amplitudes to account for the identity of the two identical $\pi^0$ mesons.
By exchanging $p$ and $p'$, the two $\pi^{0}$ momenta, we obtain the expressions for $\tilde{t}_{a}$, $\tilde{t}_{b}$, and $\tilde{t}_{c}$,
\begin{align} 
\tilde{t}_{a} &= A'\, (\vec{p}\,' + \vec{p}_{\pi^{+}})\cdot (\vec{p}_{\bar K^{0}} - \vec{p}\,) \nonumber\\
&~~~~\times\frac{1}{(p + p_{\bar K^{0}})^{2} - M_{\bar K^{*0}}^{2} + i M_{\bar K^{*0}} \Gamma_{\bar K^{*0}}}, \\
\tilde{t}_{b} &= B'\, (\vec{p}\,' + \vec{p}_{\bar K^{0}})\cdot (\vec{p}_{\pi^{+}} - \vec{p}\,) \nonumber\\
&~~~~\times\frac{1}{(p + p_{\pi^{+}})^{2} - M_{\rho^{+}}^{2} + i M_{\rho^{+}} \Gamma_{\rho^{+}}}, \\
\tilde{t}_{c} &= C'\, M_{\bar K^{*0}}^{2} (\vec{p}_{\pi^{+}} - \vec{p}\,')\cdot (\vec{p}_{\bar K^{0}} - \vec{p}\,) \nonumber\\
&~~~~\times\frac{1}{(p_{\pi^{+}} + p')^{2} - M_{\rho^{+}}^{2} + i M_{\rho^{+}} \Gamma_{\rho^{+}}} \nonumber\\
&~~~~\times \frac{1}{(p + p_{\bar K^{0}})^{2} - M_{\bar K^{*0}}^{2} + i M_{\bar K^{*0}} \Gamma_{\bar K^{*0}}}. 
\end{align}

At this stage, only six variables remain in Eq.~\eqref{eq:FinalGamma}. 
The three-momenta of the four daughter particles are constructed from these six variables, enabling the evaluation of the integral and seven distinct mass distributions through the Monte Carlo integration method.

We generate $10^8$ events, and in each event, six random values of these variables are generated. 
On one hand, the variables are constrained within the limits imposed by physics; 
on the other hand, the $\Theta$ functions determine the phase space.

We will also introduce a contact term to account for the production of the four pseudoscalars with no resonance production, which we call $t_{\rm con.}=D$, with $D$ a constant.
Next, following the experimental paper \cite{BESIII:2023qgj}, we include the contribution of the $a_1(1260)$ in the process $D^+ \to \bar K^0 a_1(1260)$; $a_1 \to \rho^+ \pi^0$ and $\rho^+\to \pi^+ \pi^0$.
For this, we follow strictly the formalism of the experimental paper \cite{BESIII:2023qgj}, and the details are given in Appendix \ref{sec:App}.
We call this term $t_{a_1}$, which symmetrized 
for the two identical $\pi^0$ gives $t_{a_1} + \tilde{t}_{a_1}$, to which we give a complex weight, $C_{a_1} e^{i\delta}$.
	
Considering all the ingredients of our model, the full amplitude is written as
\begin{align}\label{eq:t-total}
		t  = & D + A' (t_a + \tilde{t}_a) + B' (t_b + \tilde{t}_b) + C' (t_c +  \tilde{t}_c) \nonumber\\
		& + C_{a_1} e^{i\delta} (t_{a_1} + \tilde{t}_{a_1}).
\end{align}
We have then 6 free parameters for the fit.

\section{Results}
\begin{figure*}[h]
	\centering
	\includegraphics[width=0.9\textwidth]{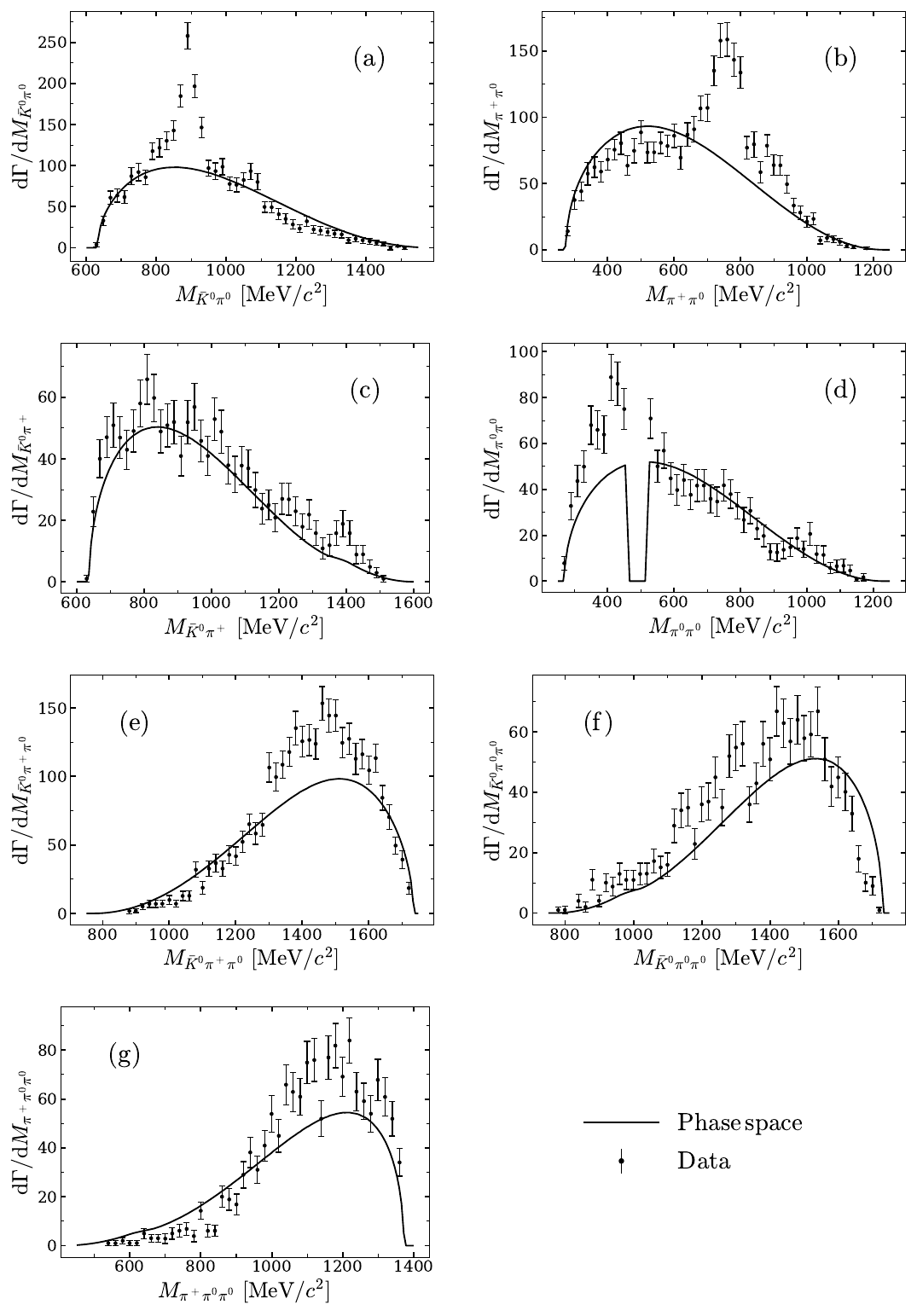}
	\vspace{-0.5cm}
	\caption{The invariant mass distributions of $\bar{K}^0\pi^0$ (a), $\pi^+\pi^0$ (b), $\bar{K}^0\pi^+$ (c), $\pi^0\pi^0$ (d), $\bar{K}^0\pi^+\pi^0$ (e), $\bar{K}^0\pi^0\pi^0$ (f) and $\pi^+\pi^0\pi^0$ (g) in the $D^+ \to \bar{K}^0\pi^+\pi^0\pi^0$ decay, taking the amplitude $t=0.1$. The black points with error bars represent experimental data from BESIII \cite{BESIII:2023qgj}. The black lines represent the phase space. The hole in figure (d) corresponds to no data taken in the experiment in that region (see text).
	}
	\label{fig:6}
\end{figure*}

At the beginning, we substitute $t = D$ in Eq.~\eqref{eq:FinalGamma} to obtain pure phase space, and the results are shown in Fig.~\ref{fig:6} along with the experimental data. The area below the curve is chosen to be close to the data for every comparison. It is evident that the phase space does not exhibit the peak structures that are clearly observed in the experimental data: the \(\bar{K}^*\) peak in the \(M_{\rm inv}(\bar{K}^0 \pi^0)\) distribution and the \(\rho^+\) peak in the \(M_{\rm inv}(\pi^+\pi^0)\) distribution.
\begin{figure*}[h]
	\includegraphics[width=0.9\textwidth]{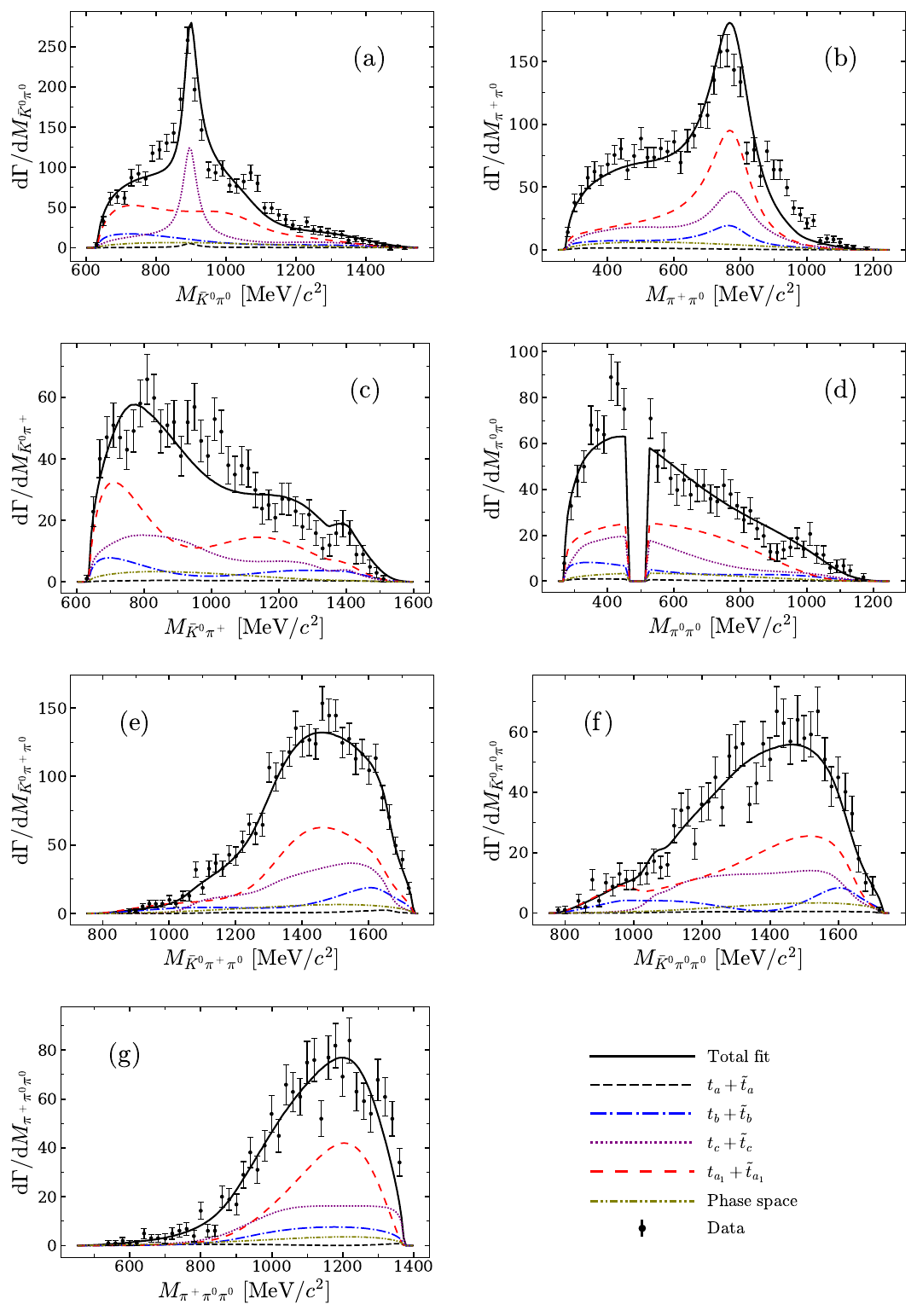}
	\vspace{-0.5cm}
	\centering
	\caption{Fitting to the data of the $\bar{K}^0 \pi^0 (a), \pi^+\pi^0 (b), \bar{K}^0 \pi^+$ (c), $\pi^0 \pi^0$ (d), $\bar{K}^0 \pi^+\pi^0$ (e), $\bar{K}^0 \pi^0 \pi^0$ (f), and $\pi^+\pi^0 \pi^0$ (g) invariant mass distributions in the $D^+ \to \bar{K}^0\pi^+\pi^0\pi^0$ decay, with the black points with error being experimental data from BESIII \cite{BESIII:2023qgj}, and the black lines the theoretical results.}
	\label{fig:7}
\end{figure*}

By using the total amplitude of Eq.~\eqref{eq:t-total}, we show the results of the global fit to the experimental data \cite{BESIII:2023qgj} in Fig.~\ref{fig:7}, which is obtained with the fit parameters values
\begin{equation}
\begin{aligned}
A' &= 0.0041, \quad B' = 0.0238, \quad C' = -0.0041, \\
C_{a1} &= -23545, \quad \delta = -1.938, \quad D = -0.0257,
\end{aligned}
\end{equation}
with a $\chi^2/{\rm d.o.f.} \simeq 1.9$.

In Fig.~\ref{fig:7}(a) for the $\bar{K}^0 \pi^0$ mass distribution, we see that the $\bar{K}^{*0}$ peak is well reproduced from the $\bar{K}^{*0}\rho^+$ production mode, but we have a non-negligible contribution from other terms.
In fact, it is remarkable to see that the lower part of the spectrum is well reproduced. 
One might be tempted to think that this is the contribution from the $K_0^*(700)$ resonance, but this is not the case since this resonance is not present in our approach. 
Instead, the largest contribution in that region comes from the $\bar{K}^0 a_1^+(1260)$ production, with posterior decay of $a_1^+(1260)$ to $\rho^+ \pi^0$.
	
In Fig.~\ref{fig:7}(b) for the $\pi^+ \pi^0$ mass distribution, we also see that the whole spectrum is fairly well reproduced, in particular the $\rho^+$ peak.
It is curious to see that the strength of the $\rho^+$ peak comes from three production mechanisms, $\bar{K}^0 a_1(1260)$ ($a_1 \to \rho^+ \pi^0$), $\bar{K}^{*0} \rho^+$ and $\bar{K}^0 \rho^+ \pi^0$, in that order of relevance.
One is called the attention to the broad bump contribution at lower energies.
It comes from a constructive interference of many mechanisms, as one can see from the figure.
	
Next we look at the $\bar{K}^0 \pi^+$ distribution in Fig.~\ref{fig:7}(c).
The spectrum is rather structureless because there is no $\bar{K}^*$ resonance with positive charge.
Once again, we obtain a fair description of the data, something not trivial in view of the many terms in our amplitude and the interference between the terms.
The bump at low invariant masses is mostly due to the $\bar{K}^0 a_1(1260)$ production mode.
	
In Fig.~\ref{fig:7}(d) we show the results of $\pi^0 \pi^0$ mass distributions.
We implement the same cut $460\mev < M_{\pi^0\pi^0} < 520\mev$, as in the experiment.
The mass distribution also does not have any particular structure and is well reproduced by our model.
	
In Fig.~\ref{fig:7}(e) we show the results for the three-body mass distribution of $\bar{K}^0 \pi^+ \pi^0$. 
The data are very well reproduced and the largest contribution, peaking at the mass of the experiment, comes from the $\bar{K}^0 a_1(1260)$ ($a_1 \to \rho^+ \pi^0$) mechanism.
	
In Fig.~\ref{fig:7}(f) the $\bar{K}^0 \pi^0 \pi^0$ mass distribution is shown, and it peaks around $1500\mev$. It is well reproduced and, once again, the $\bar{K}^0 a_1(1260)$ production is mostly responsible for the shape and strength of the distribution.
	
In Fig.~\ref{fig:7}(g) we show the mass distribution for $\pi^+ \pi^0 \pi^0$. In this case the $\bar{K}^0 a_1(1260)$ ($a_1 \to \rho^+ \pi^0$) mechanism is mostly responsible for the shape and strength of the mass distribution, which exhibits a clear broad peak in the region of the $a_1(1260)$.

\section{Conclusions}

We study theoretically the $D^+ \to K^0_S\pi^+\pi^0\pi^0$ decay process, considering various intermediate particles that decay further to produce the four daughter particles.
We begin by considering the weak decay of $D^+$ into quarks, and we examine both external and internal emission mechanisms, incorporating hadronization when necessary to generate the mesons in the final state. 
Consequently, three types of mechanisms are analyzed, and by exchanging the two neutral pions, three additional mechanisms are derived, ultimately yielding an amplitude described by only three parameters. 
To this amplitude we add a small background and the contribution of the $D^+\to \bar K^0 a_1(1260)$ with $a_1 \to \rho^+ \pi^0$ and the whole amplitude is symmetrized for the precence of two identical $\pi^0$, with 6 free parameters in total.
From that amplitude, 
a global fit to the experimental mass distributions in carried out, 
and seven mass distributions are obtained, allowing a direct comparison with the corresponding experimental data.

The mass distributions clearly show peaks corresponding to the $\rho^+$ and $\bar{K}^*$ resonances, which align well with the experimental data. 
The remaining mass distributions are also very well reproduced and we can see that the $D^+ \to \bar K^0 a_1(1260)^+, \; a_1 \to \rho^+ \pi^0$ mechanism is the dominant one in the whole process and is responsible for the line shapes of the different distributions.

\section*{Acknowledgments}
This work is partly supported by the National Natural Science Foundation of China (NSFC) under Grants No. 12575081 and No. 12365019,
and by the Natural Science Foundation of Guangxi province under Grant No. 2023JJA110076,
the Central Government Guidance Funds for Local Scientific and Technological Development, China (No. Guike ZY22096024),
the Scientific Research Foundation of Hunan Provincial Education Department under Grant No. 24B0063, and the Youth Talent Support Program of Hunan Normal University under Grant No. 2024QNTJ14.
This work is also partly supported by the Spanish Ministerio de Economia y Competitividad (MINECO) and European FEDER funds under Contracts No. FIS2017-84038-C2-1-PB, PID2020-112777GB-I00, and by Generalitat Valenciana under contract PROMETEO/2020/023. 
This project has received funding from the European Union Horizon 2020 research and innovation program under the program H2020-INFRAIA-2018-1, grant agreement No. 824093 of the STRONG-2020 project.

\appendix
\section{Contribution from the $a_1(1260)$ resonance}\label{sec:App}
According to the experimental paper \cite{BESIII:2023qgj}, in addition to the $K^*$ and $\rho$ resonances, the most significant contribution in the fitting is the $a_1(1260)$ production, with the decay process given by
\begin{equation}\label{eq:A1}
		D^+ \to \bar{K}^0 a_1(1260)^+(a_1 \to \rho^+ \pi^0) \to \bar{K}^0 \pi^+ \pi^0 \pi^0.
\end{equation}
We call the corresponding amplitude $t_{a_1}$, which is described in Ref.~\cite{BESIII:2023qgj} using covariant tensors as follows
\begin{equation}\label{eq:A2}
		t_{a_1}(p) = P_{a_1} P_{\rho^+} S X_{a_1} X_{\rho^+} X_{D^+}.
\end{equation}
The propagators of the two resonances, which describe the corresponding lineshapes, are indicated by $P$,
\begin{equation}\label{eq:A3}
\begin{aligned}
		P_{a_1} & = \frac{1}{(p_{\pi^+} + p_{\pi^0} + p_{\pi^0}')^2 - M_{a_1}^2 + iM_{a_1}\Gamma_{a_1}}, \\
		P_{\rho^+} & = \frac{1}{(p_{\pi^+} + p_{\pi^0})^2 - M_{\rho^+}^2 + iM_{\rho^+}\Gamma_{\rho^+}}.
\end{aligned}
\end{equation}
	
For a two-body decay, $a \to b+c$, we use the notation $p_a,\,p_b$ and $p_c$ for the momenta of particles $a, b$ and $c$ respectively, and let $r_a = p_a - p_b$.
The spin factor $S(p)$ in the cascade decay $D^+ \to A P_1, \;A \to V P_2$ is constructed from the spin projection operators and pure orbital angular-momentum covariant tensors as follows
\begin{equation}\label{eq:A4}
		S(p) = \tilde{t}^{(L=1)\mu}(D^+) \; P_{\mu\nu}^{(1)}(A)\; \tilde{t}^{(L=11)\nu}(V),
\end{equation}
where the spin projection operators are defined as
\begin{equation}\label{eq:A5}
		P_{\mu\nu}^{(1)}(a) = -g_{\mu\nu} + \frac{p_{a,\mu}\;p_{a,\nu}}{p_a^2}, \quad (P\text{-wave}),
\end{equation}
and the covariant tensors are given by \cite{Zou:2002ar}
\begin{equation}\label{eq:A6}
		\tilde{t}_\mu^{(L=1)}(a) = -P_{\mu\nu}^{(1)}(a)\;r_a^\nu, \quad (P\text{-wave}).
\end{equation}
	
In Eq.~\eqref{eq:A2}, the $X_{a_1, \rho^+} (X_{D^+})$ are the Blatt-Weisskopf barrier penetration factors for the intermediate resonances (the $D^+$ meson),
\begin{equation}\label{eq:A7}
\begin{aligned}
		X_{a_1}^{L=0} &= 1,  \\
		X_{\rho^+}^{L=1}(q) &= \sqrt{\frac{(q_0R)^2+1}{(qR)^2+1}}, \\
		X_{D^+}^{L=1}(q) &= \sqrt{\frac{(q_0R)^2+1}{(qR)^2+1}},
\end{aligned}
\end{equation}
where $q$ is the momentum of the final-state particle $b$ or $c$ in the rest frame of $a$,
\begin{equation}
		q(\sqrt{s}) = \frac{\sqrt{[m_a^2-(m_b+m_c)^2][m_a^2-(m_b-m_c)^2]}}{2m_a},
\end{equation}
with $m_a$, $m_b$ and $m_c$ being the invariant masses of the corresponding particles.
The value of $q_0$ is that of $q$ when taking $m_i$ as the physical mass of particle $i$.

\bibliographystyle{a}
\bibliography{refs}
\end{document}